Replace this line with article main subject (see Index Terms below)_______________________

# Exchange-Biased multi-ring Planar Hall Magnetoresistive Sensors with nT resolution in Non-Shielded Environments

J. Schmidtpeter[1,2], Proloy T. Das[1,*], Y. Zabila[1], C. Schubert[1], T. Gundrum[2], T. Wondrak[2] and D. Makarov[1,*]

[1] Institute of Ion Beam Physics and Materials Research, Helmholtz -Zentrum Dresden-Rossendorf e.V., 01328 Dresden, Germany
[2] Institute of Fluid Dynamics, Helmholtz-Zentrum Dresden-Rossendorf e. V., 01328 Dresden, Germany
* Author to whom correspondence should be addressed.

Abstract— Planar Hall magnetoresistive sensors (PHMR) are promising candidates for various magnetic sensing applications due to their high sensitivity, low power consumption, and compatibility with integrated circuit technology. However, their performance is often limited by inherent noise sources, impacting their resolution and overall sensitivity. Here the effect of three bilayer structures NiFe(10 nm)/IrMn(10 nm), NiFe(30 nm)/IrMn(10 nm), and NiFe(30 nm)/IrMn(20 nm) on noise levels is investigated at low-frequency (DC - 25 Hz). This study includes a detailed investigation on the optimization process and noise characteristics of multiring PHMR sensors, focusing on identifying and quantifying the dominant noise sources. The experimental measurements are complemented by a theoretical analysis of noise sources including thermal noise, 1/f noise, intermixing and environmental noise. The best magnetic resolution is observed for the NiFe(30 nm)/IrMn(10 nm) structure, which achieves a detectivity below 1.5 nT/√Hz at 10 Hz in a non-shielded environment at room temperature. In addition, a substantial improvement in sensitivity is observed by annealing the sensors at 250 °C for 1 hour. The findings of this study contribute to a deeper understanding of noise behavior in PHMR sensors, paving the way for developing strategies to improve their performance for demanding sensing applications at low frequencies.

## I. INTRODUCTION

The demand for sensitive and miniaturized magnetic field sensors is continuously increasing, driven by applications in various fields such as the automotive industry [Treutler 2001], consumer electronics [Melzer 2015], biomedical sensing [Murzin 2020], space exploration [Brown 2012], archaeology [Wunderlich 2022], flexible electronics [Granell 2019, Nhalil 2023], nanomagnetic bead detection [Das 2021], multi-axis sensing [Das 2024]. Compared to various magnetoresistance (MR) based sensing technologies such as the anisotropic magnetoresistance (AMR), giant magnetoresistance (GMR) and tunnel magnetoresistance (TMR), planar Hall magnetoresistance (PHMR) sensors have emerged as promising candidates due to attractive features such as their high sensitivity, low power consumption, small size and compatibility with standard microfabrication processes [Lim 2022]. The working principle of PHMR relies on the planar Hall effect, which manifests as a resistance change of a ferromagnetic material when subjected to an in-plane magnetic field. This resistance change is proportional to the sine of twice the angle between the current direction and the applied magnetic field [Mor 2017].

Despite their advantages, the performance of PHMR sensors is often limited by inherent noise sources, which can mask the desired magnetic signal and degrade the sensor's resolution. Recent studies on PHMR sensors have focused on increasing their magnetic field sensitivity and resolution, either by changing the sensor material structure or its geometry. In this, exchange biased (EB) planar-Hall sensors hold a crucial role in order to enhance sensor sensitivity and operational dynamic range [Piskin 2020, Lim 2022]. Understanding the origin and characteristics of noise in PHMR sensors plays a key role in developing strategies to mitigate its impact and improve the sensor's overall performance [Elzwawy 2021, Lim 2022].

The best resolution of planar Hall sensors is reported till date ~ 5 pT/√Hz at 10 Hz [Das 2021 and Nhalil 2019] integrating magnetic flux concentrators and in magnetic shielding. Without magnetic shielding and concentrators, the reported value is ~ 24 pT/√Hz at 10 Hz [Nhalil, 2020]. For planar Hall sensors with a meander ring structure, the best reported value is in the range of 550 pT/√Hz at 100 Hz in an unshielded environment [Jeon 2021]. It is also important to mention that there are few reports in the literature showing the performance of the sensors in unshielded environments and highlighting their operational range. In the literature, mostly IrMn(10 nm)/NiFe(10 nm) structures are used as reference. For this reason, we changed the layer thickness of the sensor stack to investigate the role of the pinning effect of the exchange bias on the sensitivity and noise of the sensor.

Therefore, the aim of this work is to further improve the performance of exchange biased (EB) PHMR sensors and to investigate the noise characteristics of the sensors. In the EB mechanism, a unidirectional anisotropy is formed inside the sensor material by a magnetic coupling between FM and AFM layers at the material interface [Nogués 1999]. A focus lies on identifying and quantifying the improved sensor material structure and the dominant noise mechanisms affecting the sensor's performance in non-shielded environments. Benchmarking to the literature review [Jeon 2021], EB PHMR meander 5-ring sensors are chosen for the present study, which demonstrates few nT resolution and higher operational range. In this work, the authors investigated the performance of different ring sensors in detail with an optimized excitation current. In summary, Jeon *et al.* reported that 5-ring sensors could be a great choice especially for extremely-low frequency measurements. Thus, based on our future application requirements (DC-150 Hz) using these sensors, here we set the sensor ring numbers to five. Furthermore, very few systematic studies are available on sensor characterization by changing the AFM/FM composition ratio, which manipulates the EB pinning strength. Therefore, here we evaluate the resolution of the different PHMR sensors based on their measured noise level and discuss their potential applicability in practical applications. In addition, this research provides insights into the noise behaviour of PHMR sensors, contributing to the development of low-noise sensor designs and signal processing techniques for enhancing their performance in various magnetic sensing applications.

*Corresponding author: PTD (p.das@hzdr.de) & DM (d.makarov@hzdr.de).

## II. PRINCIPLES OF PHMR SENSORS

PHMR sensors typically consist of a thin ferromagnetic film patterned into a cross-shaped structure [Lim 2022, Grosz 2016]. When a current is passed through the film, and an external in-plane magnetic field is applied, the resistance of the film changes due to the AMR effect. The AMR effect arises from the spin-orbit interaction, which causes the resistivity of the material to depend on the angle between the magnetization direction and the current flow direction. If current passes through the electrodes along the x axis ($I_x$) then a voltage is induced at the electrodes along both the x and y axes. For cross-type junction the effective induced planar-Hall voltages can be expressed as,

$$V_y = \frac{1}{t_{FM}} \rho_{xy}(\phi) I_x \qquad [\text{with}, \rho_{xy} = \frac{1}{2} \Delta\rho \sin(2\phi)] \quad (1)$$

Where, $V_y$ is the planar-Hall voltage, which is caused by the off-diagonal resistivity component of $\rho_{xy}$ [Lim 2022]. $\Delta\rho = \rho_{||} - \rho_{\perp}$. $\rho_{||}$ and $\rho_{\perp}$ are the resistivities parallel and perpendicular to the magnetization, respectively. $\phi$ is the angle between the bias current and the magnetization. The dimensions of the cross-type sensor are: l: length, w: width, $t_{FM}$: thickness of ferromagnetic material. In the present study, Wheatstone-bridge structured multi-ring PHMR (mPHMR) sensors are investigated. For this hybrid structure, the effective PHMR voltage can be expressed as [Lim 2022],

$$V_y = V_{offset} + V_s(\phi) \qquad (2)$$

$$\text{with,} \begin{cases} V_{offset} = \frac{R_1 R_3 - R_2 R_4}{R_1 + R_2 + R_3 + R_4} I_x \\ V_s(\phi) = \frac{1}{2} \frac{r}{w t_{FM}} \Delta\rho \sin(2\phi) I_x \end{cases}$$

Here the offset voltage is contributed by an unbalance in the arms $R_1$ -$R_4$. The desired planar-Hall magnetoresistive voltage is additionally proportional to the width and the radius of the ring, $r$.

## III. EXPERIMENTAL DETAILS

### A. Sensor Fabrication

In this study, three different bi-layered multiring PHMR (bi-mPHMR) sensor structures are patterned using a standard microfabrication process. Based on previous studies, the ring number is set to five [Jeon 2021]. The sensor width (w) is fixed to 40 µm with a gap of 0.16 mm between two consecutive rings. Thus, the total sensing area is approximately 88.94 µm². In this study, unidirectional anisotropy is developed in the sensor using the $Ni_{80}Fe_{20}$ (Py)/$Ir_{80}Mn_{20}$ bilayer EB mechanism. The full sensor stack for the bi-mPHMR is Ta (5 nm)/ Py(10-30 nm)/IrMn (10-20 nm)/Ta (5 nm) and is grown on a Si + $SiO_2$ (500 nm) wafer substrate. Here, Ta (5 nm) serves as a capping and seed layer in order to prevent oxidation and to promote better crystallinity. A BESTEC magnetron sputtering system is used in order to deposit different metal layers with a base pressure of 6 x $10^{-8}$ mbar and a process pressure (during sputtering) of 8 x $10^{-4}$ mbar in an Argon gas environment (gas flow rate: 15 sccm). The deposition rates for Ta, Py, and IrMn were 0.93 nm/s, 0.1 nm/s, and 0.2 nm/s, respectively. Note that, except for Py, all other materials are deposited in rf-sputtering mode. Py is deposited in dc-sputtering mode. An external in-plane field of 100 mT is applied to induce anisotropy in the sensor during growth. For this purpose, a standard NdFeB permanent magnet is used, which is inserted into a custom designed sample holder. The active sensing element and gold contact pads are patterned using photolithography and lift-off processes. The sensor and contact layers are structured in a two-stage lithographic process. To promote adhesion, TI Prime (Microchemicals GmbH) is spin coated at 3000 rpm for 30 s, followed by a baking step at 115°C for 3 min. After cooling, AZ5214E photoresist (Microchemicals GmbH) is spin coated at 3000 rpm for 30 s, followed by a baking step at 100°C for 30 s. Laser writing is then performed using a Heidelberg DWL66 Laser writer, before baking it at 115°C for 90 s and then UV-exposure is carried out for 45 s. In the next step, the sample is developed in a 1:5 solution of AZ351B (Microchemicals GmbH) and deionized water for 40 to 120 s. After development, a passivation in deionized water is performed for at least 2 minutes to stop further development. After the deposition of the sensing layer, the second lithography step is performed in a similar way as for the sensing layer. The fabricated sensor structure is shown in Fig. 1(a) and Fig. 1(b). For better understanding the three different sensor stacks of Ta (5 nm)/Py(10 nm)/IrMn (10 nm)/Ta (5 nm) , Ta(5 nm)/ Py(30 nm)/IrMn (10 nm)/Ta (5 nm), and Ta(5 nm)/ Py(30 nm)/IrMn (20 nm)/Ta (5 nm) are designated as S1, S2, and S3 respectively. For all the studied sensors the on-diagonal resistance ($R_{xx}$ or $R_{yy}$) and off-diagonal counterpart ($R_{xy}$) of S1, S2 and S3 are in the range between 800 and 1300 Ohms. Detailed description of the resistances is tabulated in Table - I. The sensors are bonded to a PCB with a 25 µm Al wire using the Thin Wire Wedge-Wedge Bonder 5630 by F&S BONDTEC Semiconductor GmbH with a bond time of 42 ms, 120 digits of ultrasonic power, and a bond force of 21 grams for both gold and copper bonds.

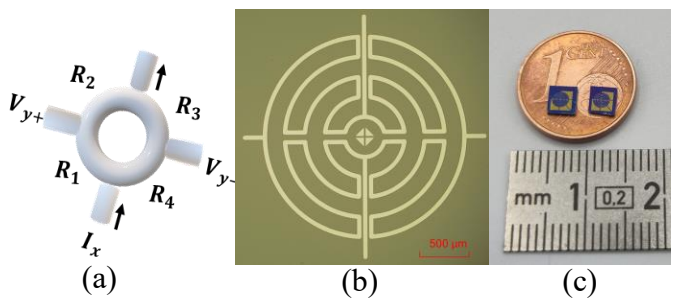

Fig. 1. *(a) Schematic of a single ring structure, (b) A geometrical design of fabricated 5-ring PHMR sensor.* (c) Qualitative size comparison between a sensor prototype and a Euro cent.

### B. Sensitivity and Noise Measurement Setup

The experimental diagram for the sensitivity measurement is shown in Fig. 3. In the first step a sinusoidal signal with a frequency of 400 Hz and an amplitude of 1 mA (in current biased voltage mode) is passed through the sensor using LTT24 (precision measuring device for data acquisition from Labortechnik Tasler GmbH) to serve as a modulation signal. In the second step, an external in-plane rotating magnetic field is generated by two nested pairs of Helmholtz coils [Podaru 2015]. The two coil pairs are arranged perpendicular to each other (as shown in Fig. 2) and are fed with a sinusoidal signal that generates a magnetic field of equal amplitude. To achieve this, the output of a two-channel function generator AFG3102 by Tektronixwas adjusted to give the corresponding current to both coils. The current was measured using the Tektronix TCP312A current probe measurement system. The reference magnetic field generated by nested Helmholtz coils were calibrated using Fluxgate sensor (model: "Magnetomat 1.782") for "Institut Dr. Foerster GmbH & Co. KG". Figure 3(a) shows the magnetic field created by both coils resulting in the change of the angle of the magnetic field shown in Fig. 3(b). In the next step, the sensor output voltage is recorded with the 24-bit ADC LTT24 data acquisition system at 100 ksamples/s. After recording, the sensor output is demodulated using a Lomb-Scargle periodogram, returning the magnetic field measurement of the sensor (see Fig.3(c)) [Ratajczack 2020]. The reference magnetic field is also recorded with the LTT24 device. Figure 3(d) exhibits the angular dependence of the output polarity of the sensor. Finally, the sensitivity of the sensor is calculated by dividing the amplitude of the sensor output ($V_y$) by the measured amplitude of the applied magnetic field ($B_{ext}$) and can additionally be mapped along the angle of the magnetic field. For all measurements, $S_y$ is determined at 1 mA. The mathematical expression for the sensor sensitivity can be expressed as follows:

$$S_y = \frac{\Delta V_y (B,T)}{\Delta B_{ext}} \qquad (3)$$

where $\Delta V_y$ is the extrapolated peak voltage related with maximum voltage, and $\Delta B_{ext}$ is the field interval from zero to maximum peak voltage field. Note that the applied magnetic field is eventually proportional to the total system anisotropy [Grosz 2016, Lim 2022]. Note that, for a m-ring PHMR sensor, the sensor sensitivity is the individual sum of each ring structure. In exchange bias structures, magnetic anisotropy is induced due to the exchange coupling field, $H_{ex}$ and the induced anisotropy field, $H_a$. Thus, total local field anisotropy is expressed as [Jeon 2021 and Sinha 2012],

$$\Delta H_{ext} = \sqrt{\frac{2}{3}} \left( H_{ex}(T) + H_a(T) \right) \qquad (4)$$

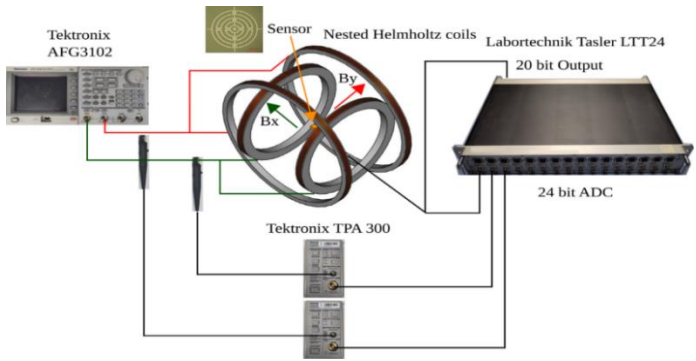

Fig. 2. Schematic of the sensitivity measurement set-up. Similar setup is used for noise measurements.

The sensor noise measurements were carried out in a non-shielded environment using conventional AC modulation technique without any magnetic shielding. By exciting the sensor with an alternating current, its output signal and intrinsic *1/f* noise are transferred to frequencies at which the *1/f* noise of the preamplifier and the electronic system noise can be neglected. In order to perform these measurements a Tensormeter Model RTM 2 (HZDR Innovation

GmbH) with an inbuilt preamplifier (down to 2.4 nV/√Hz) and the output data is recorded using an ADC of the same system in a differential mode. The sensor biasing was carried out with a low-noise Tensormeter current source and an optimized current amplitude was set to 1 mA. The measured signal is modulated at 1025 Hz at a 1 Hz resolution bandwidth (BW). The noise measurements are performed in a 100 Hz measurement bandwidth. The noise measurements are performed in a 100 Hz measurement bandwidth. In this study a 0.005 s integration time for demodulation is used. The sampling rate of the data acquisition system is chosen high above the Nyquist-Shannon sampling limit. All of the measurements were carried out in dc coupling mode in order to capture all frequency responses. In order to reduce the environmental noise, and other electromagnetic interferences (EMI), the sensor is placed in close proximity to the Tensormeter.

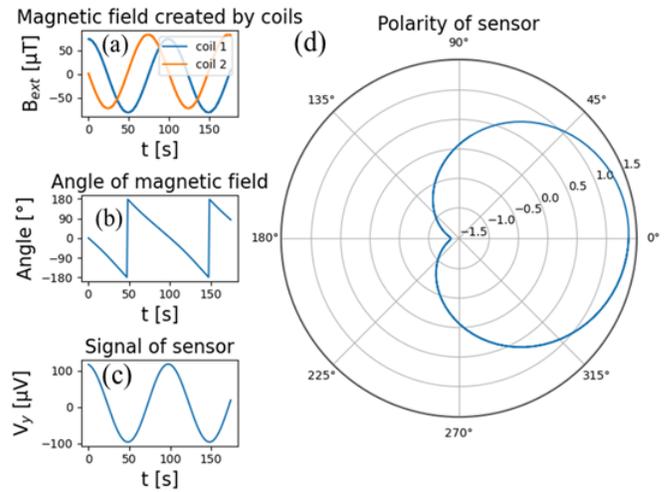

Fig. 3. (a) Magnetic field created by coils, (b) corresponding angle of magnetic field with time (s), (c) Output signal of the sensor as measured by LTT24 ADC, (d) polar plot of the sensor signal with rotation vector of the external magnetic field.

## C. Noise of PHMR sensors

The total electronic noise ($e_\Sigma$) of PHMR sensors has four major components such as low frequency $1/f$ noise ($e_{1/f}^2$), thermal or Johnson noise ($e_{thermal}^2$), preamplifier noise ($e_{amp}^2$) and environmental noise, which can be expressed as,

$$e_\Sigma = \sqrt{e_{1/f}^2 + e_{thermal}^2 + e_{mix}^2 + e_{amp}^2 + e_{env}^2} \qquad (5)$$

The sensor $1/f$ noise which is intrinsic in nature is generally attributed to fluctuations in material properties and scattering mechanisms within the sensor's ferromagnetic layer. For PHMR sensors, it can be expressed mathematically as [Qiu 2018, Grosz 2016],

$$e_{1/f}^2 = \frac{\delta_H V_B^2}{N \cdot vol \cdot f^\beta} \qquad (6)$$

Here, $\delta_H$: Hooge's parameter, $V_B$: bias voltage, N: total carrier number, vol: sensing volume, f: frequency, and $\beta$ is close to1. As we can see from eq. (6), it is possible to minimize $1/f$ noise by employing suitable mitigation strategies such as optimization of bias current, increasing the sensing volume, etc.

Thermal noise, also known as Johnson-Nyquist noise, is an inherent noise source present in all conductive materials due to thermal motion of charge carriers. The spectral density of thermal noise is white, meaning it is constant over frequency, and can be calculated using the following equation:

$$e_{thermal}^2 = 4k_B T R_{xy} \quad (7)$$

Here, $k_B$ is the Boltzmann constant ($1.380649\times10^{-23}$ J·K$^{-1}$), T is the temperature and $R_{xy}$ is the sensor resistance along the hard axis. Note that, $R_{xy}$ has a linear dependency with the sensor thickness and width [Mor 2017].

Intermixing noise in the PHMR sensor can be written as,

$$e_{mix}^2 = V_{offset}^2 \left(\frac{\delta I_n}{I_x}\right) \quad (8)$$

where $\delta I_n$ is the Nyquist noise of the operating current from the current source. Due to a mismatch of the PHMR electrode arms, there is a voltage drop along the y-axis corresponding to the offset voltage ($V_{offset}$) of the PHMR sensor. In this case, both sensor voltage ($V_y$) and $R_{xy}$ have a non-zero value due to a fabrication error. Here $I_x$ and $R_{xy}$ are the sensor operating current and off-diagonal PHMR tensor component [Lee 2021].

In this study, the baseline of the preamplifier noise is down to 2.4 nV/√Hz and environmental noise at room temperature includes all possible electromagnetic interferences from nearby electronics, vibration etc. Thus, field resolution or detectivity of the magnetic field sensor can be defined as,

$$D(T,V) = \frac{e_{\Sigma}}{S_y} \quad \left[\frac{T}{\sqrt{Hz}}\right] \quad (9)$$

## D. Results and Discussions

### I. Sensor Characterization:

The fabricated PHMR sensors are characterized to determine their magnetic and electronic properties. It includes the measurement performance of the sensor's sensitivity, linearity range, and hysteresis. Note that, in this study the applied field, $B_y$ is the small compared to the total magnetic anisotropy, $H_a + H_{ex}$ in order to reduce the sensor offset and to increase the linearity [Piskin 2020]. The measured sensitivity for S1 is ~ 1.41 x $10^3$ V/TA whereas for S2 and S3, the sensitivities are 1.61 x $10^3$ V/TA and 1.57 x $10^3$ V/TA, respectively. Sensor linearity quantifies the deviation from a linear relationship between the sensor's output and the applied magnetic field. The linearity range for these sensors lies within an operational range of 20 mT. Only a negligible hysteresis is observed.

In order to increase the sensitivity, an annealing measurement is conducted for S2 using a box furnace (Goldbrunn 450 Vacuum Dryer). The annealing temperature was set to 250 ℃ for 1 hour at ≤20 mbar air pressure. After annealing, a 27% enhancement in sensitivity is observed for the S2 sample. The $S_y$ is increased to 2.76 x $10^3$ V/TA from 1.61 x $10^3$ V/TA. Further enhancements in sensitivity might be possible by manipulating the sensor material and annealing parameters. Note that, this sensitivity enhancement has no effect on the sensor noise level which will be discussed in detail in next section. Furthermore, few more actions (such as changing the gas, and increasing the annealing temperature and time) can be taken into consideration for further enhancement in sensitivity.

### II. Noise spectral density (NSD):

Figure 5 represents the NSD of all studied PHMR samples at 1 mA excitation current up to 25 Hz bandwidth. The NSD is fitted employing the equation [Nhalil 2019],

$$Noise = \sqrt{a_0^2 + \left(\frac{a_1}{f^{\gamma}}\right)^2} \quad \left[\frac{V}{\sqrt{Hz}}\right] \quad (10)$$

where, $a_0$, $a_1$ and $\gamma$ are the fit parameters. Here, $Noise$ stands for total noise of the sensor as stated in eq. (5). Usually, the exponent $\gamma$ remains $\leq 1$. The fit parameters and the extracted detectivity (resolution) at 0.1 Hz, and 10 Hz are summarized in Table - I. It is found that the detectivity (resolution) of S2 at 0.1 Hz (7.1 nT/√Hz) is better than that of the other two sensors. A further detailed investigation is required to explain this intriguing $1/f$ nature of S2 in the context of exchange pinning strength at the interface. Note that for S1 the $1/f$ noise might be considerably higher because of stronger exchange coupling. For all sensors the corner frequency is found to be considerably lower compared to S1, especially for S2 it is found to be below 0.5 Hz. However, the best resolution of ~ 1.8 nT/√Hz is achieved at 10 Hz for the S3, while at 0.1 Hz the detectivity ($D$) is found around 39.7 nT/√Hz. The thermal noise for the S2 is calculated around 3.8 nV/√Hz (see eq. 7) and the reported preamplifier noise is 2.4 nV/√Hz. Thus, the noise contribution from these components is 4.5 nV/√Hz. However, the total white noise for S2 is measured as ~ 5.6 nV/√Hz. The excess noise, 3.3 nV/√Hz originates in the offset contribution and environmental components. Similarly, for S1 and S3 the excess noise can be attributed to intermixing and non-shielded environmental components. From these results, it can be seen that S2 is more promising compared to the other sensors. Thus, the next step we investigated the effect of annealing on the NSD for S2 and no considerable change is observed. Finally, the best effective resolution is achieved around 1.3 nT/√Hz at 10 Hz.

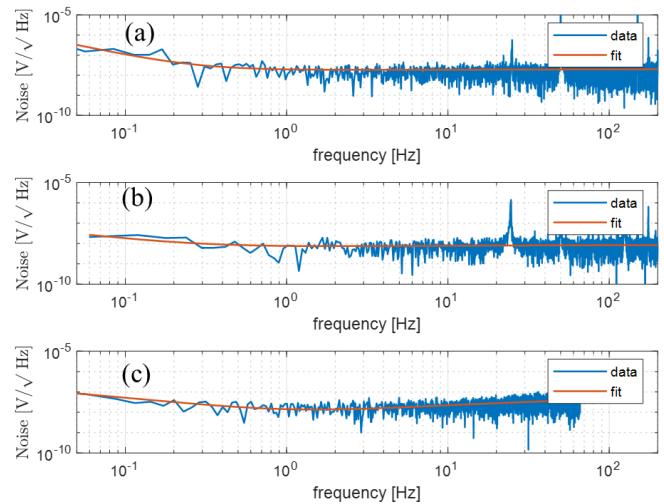

Fig. 4. Total noise versus frequency for (a) S1, (b) S2, (c) S3 sensors. For the optimum excitation current amplitude of 1 mA, both the sensor noise and the noise fit are shown.

In addition, Fig. 5 demonstrates that annealing effect has no pivotal role to suppress the NSD level of S2.

## IV CONCLUSION

In summary, we have investigated the magnetic field resolution of three m-biPHMR sensors with 5 rings. By exciting the sensor with an optimized alternating current and optimizing the sensor thickness, we achieved a 1.8 nT/√Hz resolution at 10 Hz for S3. Interestingly, the $1/f$ noise component for S2 is much less than for the other two sensors. In addition, a 27% sensitivity improvement is observed by annealing S2 at 250 ℃. Thus, after annealing the best resolution is found to be better than 1.5 nT/√Hz for S2. The operational range for the sensors is up to 20 mT. The electrical and magneto-transport characteristics of the PHMR sensors are highly reproducible over all tested samples and environments. The results indicate that these bi-layered PHMR sensors are promising for future sensor applications that require a low magnetic field sensing such as non-destructive testing (NDT) applications.

Table - I: PHMR sensor parameters

| Sensor name | $R_{yx}$ (Ohm) | $R_{yy}$ (Ohm) | $S_y$ (V/TA) x $10^3$ | Fit parameters from noise model eq. (10) | | | Noise (nV/√Hz) | | D (nT/√Hz) | |
|---|---|---|---|---|---|---|---|---|---|---|
| | | | | $a_0$ | $a_1$ | $\gamma$ | at 0.1 Hz | at 10 Hz | at 0.1 Hz | at 10 Hz |
| S1 | 2950 | 4120 | 1.42 | $9.8 \times 10^{-9}$ | $1.5 \times 10^{-8}$ | 0.86 | 111.9 | 9.8 | 78.8 | 6.9 |
| S2 | 860 | 1048 | 1.61 | $5.5 \times 10^{-9}$ | $1.1 \times 10^{-9}$ | 0.94 | 11.5 | 5.6 | 7.1 | 3.5 |
| S3 | 525 | 610 | 1.57 | $2.9 \times 10^{-9}$ | $1.2 \times 10^{-8}$ | 0.72 | 62.3 | 2.9 | 39.7 | 1.8 |

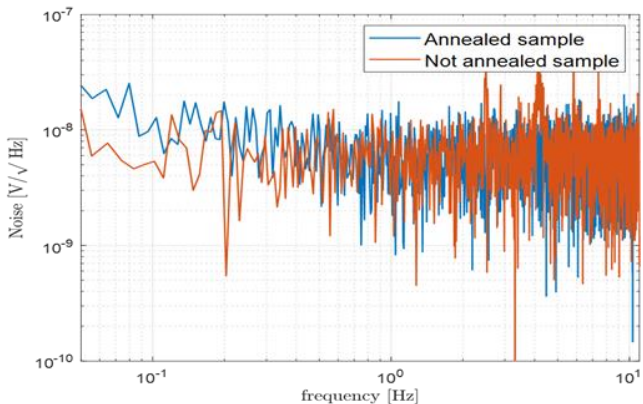

Fig. 5. NSD of S2 pre-annealed and post-annealed at an excitation current of 1 mA.

## ACKNOWLEDGMENT

This work was supported in part by European Commission's HORIZON-TMA-MSCA-PF-EF under the Marie Skłodowska-Curie grant agreement No. 101106524, ERC grant 3DmultiFerro (Project number: 101141331) and German Research Foundation (project MA5144/33-1). We would also like to express our gratitude to Dr. H. Nhalil (Christian-Albrechts University of Kiel (CAU)) for his valuable and helpful suggestions during the improvement stage of PHMR sensors.